\documentclass[]{pasj02} 
\usepackage{natbib} 
\usepackage{url} 
\usepackage[switch,mathlines]{lineno} 
\usepackage{subcaption}
\usepackage{xcolor}

\jyear{2024}
\Received{}
\Accepted{}

\begin{document} 

\title{A CO detection in the off-plane region of the edge-on galaxy NGC~4565 with the Nobeyama 45-m telescope}

\author{
 Ren \textsc{Matsusaka},\altaffilmark{1}\altemailmark\orcid{0000-0001-5509-8218} \email{ren.matsusaka.jp@gmail.com} 
 Fumi \textsc{Egusa},\altaffilmark{1}\orcid{0000-0002-1639-1515}
 Fumiya \textsc{Maeda}\altaffilmark{2}\orcid{0000-0002-8868-1255}
 and Toshihiro \textsc{Handa}\altaffilmark{3} 
}
\altaffiltext{1}{Institute of Astronomy, Graduate School of Science, The University of Tokyo, 2-21-1 Osawa, Mitaka, Tokyo 181-0015, Japan}
\altaffiltext{2}{Research Center for Physics and Mathematics, Osaka Electro-Communication University, 18-8 Hatsucho, Neyagawa, Osaka, 572-8530, Japan}
\altaffiltext{3}{Division of Liberal Arts, Kogakuin University, 2665-1 Nakano-cho, Hachioji, Tokyo 192-0015, Japan}



\maketitle

\begin{abstract}
Understanding the cycling of interstellar medium (ISM) between the galactic plane and off-plane regions is crucial for tracing the evolution of disk galaxies. We present $^{12}$CO($J$ = 1--0) multi-pointing observations of the edge-on, Milky Way–like galaxy NGC 4565 obtained with the Nobeyama 45-m telescope (angular resolution 14\arcsec $\sim 0.8$ kpc). Along a prominent dust filament, we detect significant CO emission at three off-plane positions located above the galactic plane. After evaluating possible beam-pattern contamination, the detections remain robust. The derived per-beam molecular masses are $M_{\rm mol}\simeq(2.1$--$4.3)\times10^{7}\,M_\odot$. While the off-plane spectra show broader effective line widths ($\sigma_{\rm eff}=83$--115 km s$^{-1}$) than the disk spectra, they contain CO components consistent with the local disk rotation. The mean observed off-plane CO intensity fraction is about 0.34. Comparison with geometrically thin-disk models suggests that this large fraction is best explained by gas above the disk. The large $\sigma_{\rm eff}$ values are partly attributable to a high-velocity component with molecular gas mass $M_{\rm mol}^{\rm HV}\sim 10^7\,M_\odot$ that is offset by $\sim 100~\rm km~s^{-1}$ from the local disk velocity. The kinetic energy of this component is estimated to be $E_{\rm kin}\sim10^{54}$ erg. Such a large energy requirement is difficult to explain by disk-driven feedback in NGC~4565, which has a Milky-Way-like star formation rate. External inflow is therefore one possibility.
\end{abstract}

\section{Introduction}
The evolution of galaxies is strongly regulated by the cycling of interstellar medium (ISM) between galaxies and their surrounding environments \citep[e.g.][]{LillyETAL2013}. In particular, vertical gas exchange across the galactic plane is a key process, as it directly influences the replenishment of the raw material for star formation and the regulation of the disk–halo ISM cycle. The outflows and fountains have been extensively studied primarily using ionized and atomic tracers \citep[e.g.,][]{Rossa&Dettmar2000,SancisiETAL2008,HayakawaETAL2024}. Most observations of off-plane molecular gas have been concentrated on environments around active galactic nuclei (AGN) or in starburst galaxies \citep[e.g.,][]{LeroyETAL2015,WalterETAL2017,BaoETAL2024}. Therefore, in normal (Milky Way–like) galaxies, the distribution and physical conditions of off-plane molecular gas remain poorly constrained.

A large fraction of molecular gas may exist outside the thin disk, potentially playing an important role in the disk--halo ISM cycle. For example, CO line-width analyses of face-on nearby galaxies suggest vertically extended, low-density molecular components beyond the thin disk \citep{Caldu-PrimoETAL2013}. Furthermore interferometric surveys indicate that such thick molecular gas could account for up to $\sim50\%$ of the total molecular mass \citep{PetyETAL2013}. High-sensitivity, high-resolution observations of the face-on galaxy M83 have identified molecular high-velocity clouds (HVCs) that may trace off-plane gas based on their line-of-sight velocity offsets from disk rotation \citep{NagataETAL2025}.

While several pioneering studies in the 1990s reported molecular emission extending up to $\sim$1 kpc above the galactic plane in the edge-on galaxy NGC~891 \citep[e.g.,][]{Garcia-BurilloETAL1992}, robust detections of off-plane molecular gas in normal disk galaxies remain limited. Recently, \citet{Jimenez-LopezETAL2026arXiv} revisited the disk--halo interface of NGC~891 with IRAM 30-m $^{12}$CO($J$=2--1) mapping and reported a vertically extended molecular component. This result reinforces the idea that molecular gas can be present at kpc-scale heights above the disk even in non-starburst edge-on galaxies. For NGC~4565, another nearby edge-on galaxy, $^{12}$CO($J$=1--0) observations with the Nobeyama 45-m telescope were presented by \citet{Sofue&Nakai1994}. Their data showed faint off-plane emission, but the peak brightness temperature appears to be below the $3\sigma$ level, and possible beam-pattern contamination was not considered.

Complementary evidence comes from dust observations. A few dust filaments oriented nearly perpendicular to the disk of NGC~4565 have been noted in previous work \citep[as reviewed by][]{Dahlem1997}. In the \textit{HST} F450W/F814W images used here, originally obtained for the globular-cluster study of \citet{KisslerETAL1999}, we identify vertically extended dust features (see Figure~\ref{fig:fov}). Given the close association between dust and molecular gas, molecular gas may exist at similar heights in the disk--halo interface. To directly test this possibility, molecular line observations are needed because they probe both the vertical structure and line-of-sight kinematics of molecular gas.

We therefore targeted the Milky-Way-like, nearby edge-on galaxy NGC~4565 (Table~\ref{tab:ngc4565_parameters}). Deep optical imaging shows no prominent tidal features, suggesting that NGC~4565 has not experienced a recent major merger or other strong interaction \citep{MosenkovETAL2020}. Motivated by the vertically extended dust features in our observed field, we carried out high-sensitivity multi-pointing $^{12}$CO($J$=1--0) position-switching observations toward this field with the Nobeyama 45-m telescope to investigate the galaxy's off-plane molecular component.

\begin{table}
    \caption{Physical and Observational parameters of NGC~4565}
    \begin{tabular}{ll}
        \hline 
        \hline
        Basic parameters\\
        \hline
        \multicolumn{2}{l}{The centre position ($J$2000)} \\
        \quad R.A. & 12$^{\mathrm{h}}$36$^{\mathrm{m}}$20$\fs$78 \\
        \quad Decl. & +25$\degree$59$'$15$\farcs$6 \\
        Position angle & 134$\degree$ \\
        Optical Radius ($R_{25}$) & 7.95\arcmin\\
        Inclination ($i$) & 88.5$\pm{0.5}\degree$ \\
        Distance ($D$)& 11.9$^{+0.3}_{-0.2}$ Mpc \\
        Systemic velocity ($V_{\mathrm{sys}}$) & 1230 ($\pm$5) km s$^{-1}$ \\
        Rotation velocity ($V_{\mathrm{rot}}$) & 250$\pm{5}$ km s$^{-1}$ \\
        Total mass ($M_{\rm tot}$) & $\sim10^{12}$ $M_{\odot}$\\
        Stellar mass ($M_\star$) & 5.65$\times10^{10}$ $M_{\odot}$\\
        Star formation rate (SFR) & 1.02 $\pm$ 0.10 $M_{\odot}$ yr$^{-1}$\\
        \hline
        Observational parameters\\
        \hline
        Observing season & March--April 2024\\
        Rest frequency & 115.2712018 GHz \\
        Angular (spatial) resolution ($\theta_{\rm mb}$) & 14\arcsec ($\sim$800 pc) \\
        Velocity resolution after binning & 10.0 km s$^{-1}$\\
        Sensitivity ($T_{\rm rms}$) & 9.1--19.5 mK\\
        Pointing calibrator & R~Leo\\
        \quad R.A. & 09$^{\mathrm{h}}$47$^{\mathrm{m}}$33$\fs$49 \\
        \quad Decl. & +11$\degree$25$'$43$\farcs$6 \\
        \hline
    \end{tabular}
    \label{tab:ngc4565_parameters}
    \begin{tabnote}
    {We adopt the inclination from \protect\citet{Martinez-LombillaETAL2019}. $V_{\mathrm{sys}}$ and rotation $V_{\mathrm{rot}}$ are taken from \protect\citet{Sofue&Nakai1994}. $M_{\rm tot}$ is taken from \protect\citet{HuchtmeierETAL1980}, while $M_\star$ and SFR are taken from \protect\citet{ZhengETAL2022}. Other values are taken from the values summarized in \citet{Radburn-SmithETAL2011}.}
    \end{tabnote}
\end{table}

In this letter, we report the detection of off-plane molecular emission and characterize its kinematic properties. Then we discuss its origin.
 
\section{Observations and data reduction}\label{sec:data}
\begin{figure}
  \includegraphics[width=\linewidth,trim={5 -2 5 2}, clip]{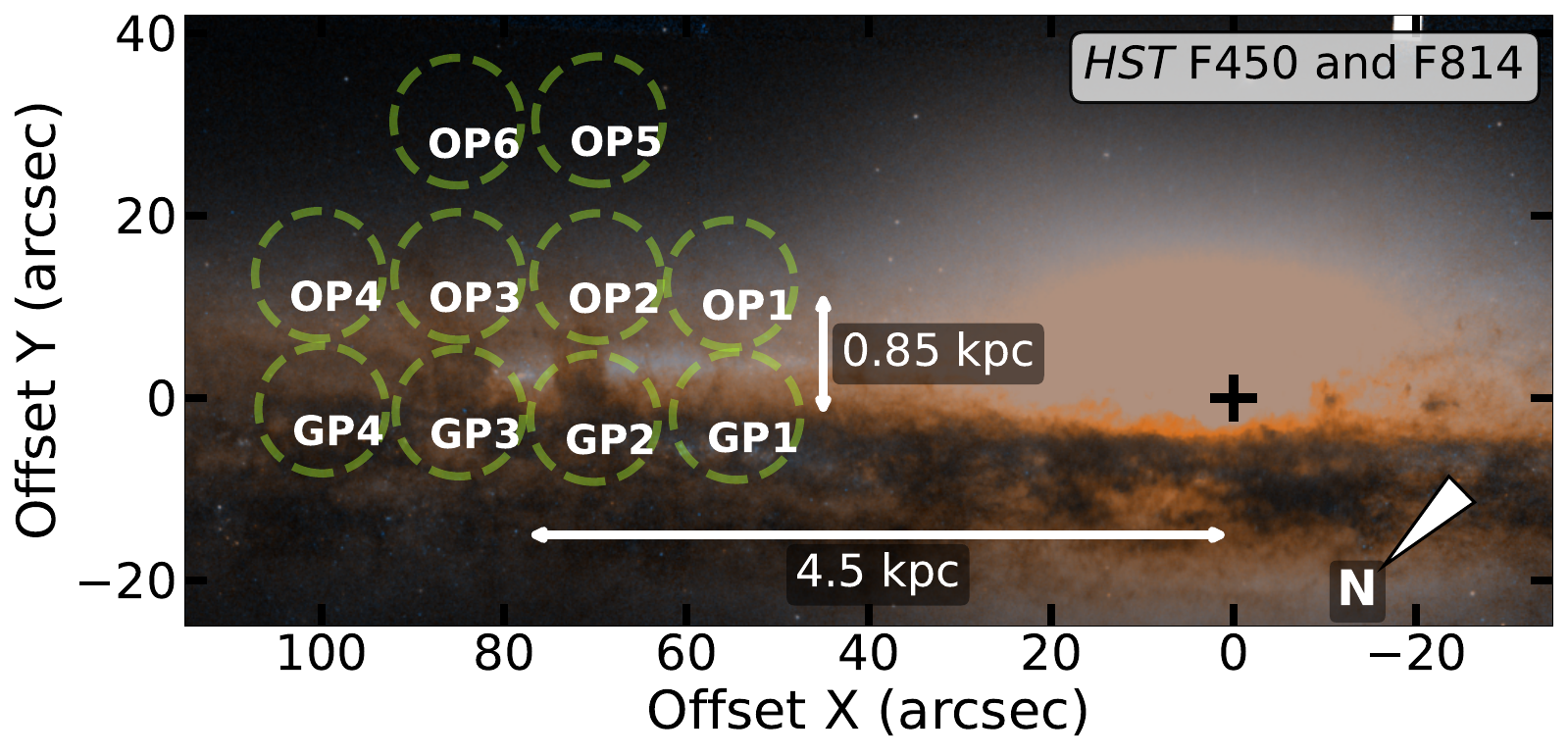} 
    \caption{Our observation points overlaid on an \protect\it{HST} \rm{} image of the dust lane region of NGC~4565. The optical image combines the F450W and F814W data. The green dashed circles represent the Nobeyama 45-m telescope beam size (14\arcsec), covering both disk positions (GP1--GP4) and off-plane positions (OP1--OP6). The black cross marks the galactic centre of NGC~4565. The north indicator in the lower-right corner shows the north direction in the R.A.--Decl. coordinate system. The horizontal and vertical arrows indicate the projected galactocentric distance and the height above the galactic plane, respectively.
    {Alt text: HST image with CO observing positions in NGC 4565.}
    }
\label{fig:fov}
\end{figure}
We conducted $^{12}$CO($J=1$--$0$) observations toward NGC~4565 using the Nobeyama 45-m telescope on 2024 March and April. The target fields consist of ten positions placed around one of the vertical dust structures seen in the \textit{HST} image\footnote{The \textit{HST} data were obtained from the Mikulski Archive for Space Telescopes (MAST), operated by the Space Telescope Science Institute; \url{https://archive.stsci.edu/}.} and the underlying mid-plane dust lane at a projected galactocentric distance of \(\sim4.5\) kpc on the sky (Figure~\ref{fig:fov}). In this study, we define the $(X,Y)$ coordinate system with its origin at the galactic centre of NGC~4565; the $X$ axis is aligned with the projected major axis at a position angle of \(134^\circ\), and the $Y$ axis is perpendicular to it. The observations were carried out in position-switching mode with FOREST (FOur-beam REceiver System on the 45-m Telescope; \citealt{MinamidaniETAL2016}), a four-beam, dual-polarization, sideband-separating SIS receiver. The specification of the observations is summarized in Table~\ref{tab:ngc4565_parameters}.

The beam size at 115\,GHz is approximately 14$^{\prime\prime}$ (in full width at half maximum: FWHM), corresponding to $\sim$0.8\,kpc at the distance of NGC~4565. The observed positions are separated by approximately one beam size and arranged to sample both the galactic plane (GP1--GP4) and off-plane (OP1--OP6) positions, as shown in Figure~\ref{fig:fov}.

The backend used was the Spectral Analysis Machine for the 45-m Telescope \citep[SAM45;][]{KamazakiETAL2012} spectrometer, an FX-type digital correlator composed of 16 arrays with 4096 channels each. Two arrays were assigned to each beam and polarization. The frequency bandwidth was set to 2~GHz with a channel spacing of 488.28~kHz, yielding a native velocity resolution of 1.3\,km\,s$^{-1}$ at 115~GHz. With a central velocity of $1230\,\mathrm{km\,s^{-1}}$ for the $^{12}$CO($J$=1--0) line, this setup covers radial velocities from approximately $-1400$ to $+3800\,\mathrm{km\,s^{-1}}$.

The integration time per scan was 10 s. The off-source reference position was set by an offset of $\rm(\Delta R.A.,\Delta Decl.)=(+10^\prime,+10^\prime)$ from each on-source position. Pointing accuracy was monitored every 40--80 minutes using nearby SiO maser sources (R~Leo), and the typical pointing error was within 3$^{\prime\prime}$. System temperatures ranged from 250 to 450\,K. The line intensity was calibrated using the standard chopper-wheel method \citep{Ulich&Haas1976}. We adopted a main-beam efficiency of $\eta_{\rm mb}=0.38$ as provided by the observatory\footnote{Status Report for Charged Telescope Time (\url{https://www.nro.nao.ac.jp/~nro45mrt/html/prop/status/Status_latest.html})}, and calculated the main-beam temperature ($T_{\rm mb}$) from the antenna temperature ($T_{\rm A}^\ast$) as $T_{\rm mb}=T_{\rm A}^\ast/\eta_{\rm mb}$.

\begin{table*}
    \caption{Observed parameters of $^{12}$CO~($J=1$--$0$) line at positions in the field}
    \begin{tabular}{ccccccccc}
        \hline 
        \hline
        pos. & $(X,Y)$ [arcsec] & $z_{\rm p}$ [kpc] 
        & $T_{\rm rms}$ [mK] & $T_{\rm pk}$ [mK] 
        & $v_{\rm pk}$ [km s$^{-1}$] 
        & $W_{\rm CO}$ [K km s$^{-1}$] 
        & $\sigma_{\rm eff}$ [km s$^{-1}$] 
        & $M_{\rm mol}$ [$10^7M_\odot$] \\
        (1)& (2) & (3) & (4) & (5) & (6) & (7) & (8)&(9)\\
        \hline
        GP1 & $(-52.7, -2.5)$ & 0.00 & 14.2 & 264.4 & 1380 & $14.5\pm0.5$ & $55.0\pm3.9$ & $4.68\pm0.16$\\
        GP2 & $(-68.1, -2.9)$ & 0.00 & 12.2 & 264.1 & 1420 & $16.9\pm0.5$ & $63.8\pm3.6$ & $5.42\pm0.16$\\
        GP3 & $(-82.8, -2.6)$ & 0.00 & 15.5 & 419.6 & 1470 & $23.7\pm0.6$ & $56.5\pm2.6$ & $7.61\pm0.19$\\
        GP4 & $(-98.2, -3.1)$ & 0.00 & 19.5 & 383.3 & 1490 & $21.9\pm0.8$ & $56.6\pm3.5$ & $7.03\pm0.26$\\
        OP1 & $(-53.5, 12.2)$ & 0.85 & 14.8 & 94.8 & 1380 & $7.8\pm0.6$ & $82.6\pm14.3$ & $2.52\pm0.19$\\
            &&&&&& $(3.3\pm0.4)$ && $(1.05\pm0.13)$\\
        OP2 & $(-68.2, 12.5)$ & 0.89 & 9.1 & 116.5 & 1410 & $13.4\pm0.4$ & $114.6\pm9.6$ & $4.29\pm0.13$\\
            &&&&&& $(6.2\pm0.3)$ && $(1.99\pm0.10)$\\
        OP3 & $(-83.6, 12.0)$ & 0.85 & 10.0 & 78.6 & 1350 & $6.6\pm0.4$ & $83.9\pm12.0$ & $2.12\pm0.13$\\
            &&&&&& $(4.6\pm0.3)$ && $(1.48\pm0.10)$\\
        OP4 & $(-99.0, 11.6)$ & 0.85 & 15.6 & 63.4 & 1470 & $2.2\pm0.4$ & $34.7\pm10.5$ & $0.71\pm0.13$\\
        OP5 & $(-68.6, 29.5)$ & 1.87 & 14.9 & $<44.7$ & \ldots & $<1.4$ & \ldots & $<0.46$\\
        OP6 & $(-84.0, 29.1)$ & 1.83 & 14.9 & 54.0 & 1430 & $1.9\pm0.4$ & $35.6\pm12.1$ & $0.62\pm0.13$\\
        \hline
    \end{tabular}
    \label{tab:obs_parameters}
    \begin{tabnote}
    (1) Position label used throughout the text and figures. 
    (2) Offset positions $(X,Y)$ in arcsec, measured relative to the galactic centre along the projected major and minor axes of NGC~4565, respectively. 
    (3) Projected vertical distance from the corresponding galactic-plane position. We use GP1, GP2, GP3, and GP4 as the reference positions for OP1, OP2, OP3, and OP4, respectively, and GP2 and GP3 for OP5 and OP6, respectively. 
    (4) $T_{\rm rms}$ noise level of the $^{12}$CO($J$=1--0) spectrum in units of mK on the main-beam temperature scale. 
    (5) Peak intensity ($T_{\rm pk}$) of the $^{12}$CO($J$=1--0) line in mK. 
    (6) $v_{\rm pk}$ is the velocity at the peak intensity. 
    (7) Integrated $^{12}$CO($J$=1--0) intensity $W_{\rm CO}$ in K km s$^{-1}$; for OP1--OP3, values in parentheses give the contribution from the observationally defined high-velocity component, i.e., emission integrated outside the blue-hatched local-disk velocity window in Figure~\ref{fig:spectrum}.
    (8) $\sigma_{\rm eff}=W_{\rm CO}/T_{\rm pk}$ is an effective line width. 
    (9) Molecular gas mass derived from $W_{\rm CO}$ using the Galactic conversion factor. The quoted uncertainties include only the statistical uncertainties propagated from $W_{\rm CO}$. For OP5, we provide 3$\sigma$ upper limits assuming a 100~km~s$^{-1}$ integration width, corresponding to ten 10~km~s$^{-1}$ channels.
    \end{tabnote}
\end{table*}

Data reduction proceeded as follows. Scans taken under high wind conditions ($\ge$5 \,m\,s$^{-1}$) were flagged, and an 11th-order polynomial baseline was subtracted from the remaining scans \citep[see also][]{KalberlaETAL2010}. In our spectra, 3rd-order fits left residual  ripples, whereas the 11th-order fits reduced the rms without significantly changing the line profiles or integrated intensities. The baseline was fitted to line-free channels within $v=500$--2000~km~s$^{-1}$, selecting appropriate ranges on both sides of the line window ($v=1300$--1500~km~s$^{-1}$) by visual inspection to minimize the impact of residual artifact features. Although this step is inherently subjective, we repeated the visual inspection and flagging process several times on separate occasions, confirming that the results were consistent and reproducible. The accepted scans were then co-added for each array, and the two orthogonal polarizations were combined to produce the final spectrum. The effective on-source integration time per position was approximately 1.0--2.0 hours. Finally, the spectra were binned to a resolution of 10\,km\,s$^{-1}$ (Figure~\ref{fig:spectrum}). The resulting root mean square ($T_{\rm rms}$) noise levels, measured on the $T_{\rm mb}$ scale, ranged from 9.1 to 19.5\,mK.

\section{Results}\label{sec:results}

\subsection{Measurements of physical quantities}
\label{analysis}
\begin{figure*}
  \includegraphics[width=\linewidth,trim={5 -5 5 5}, clip]{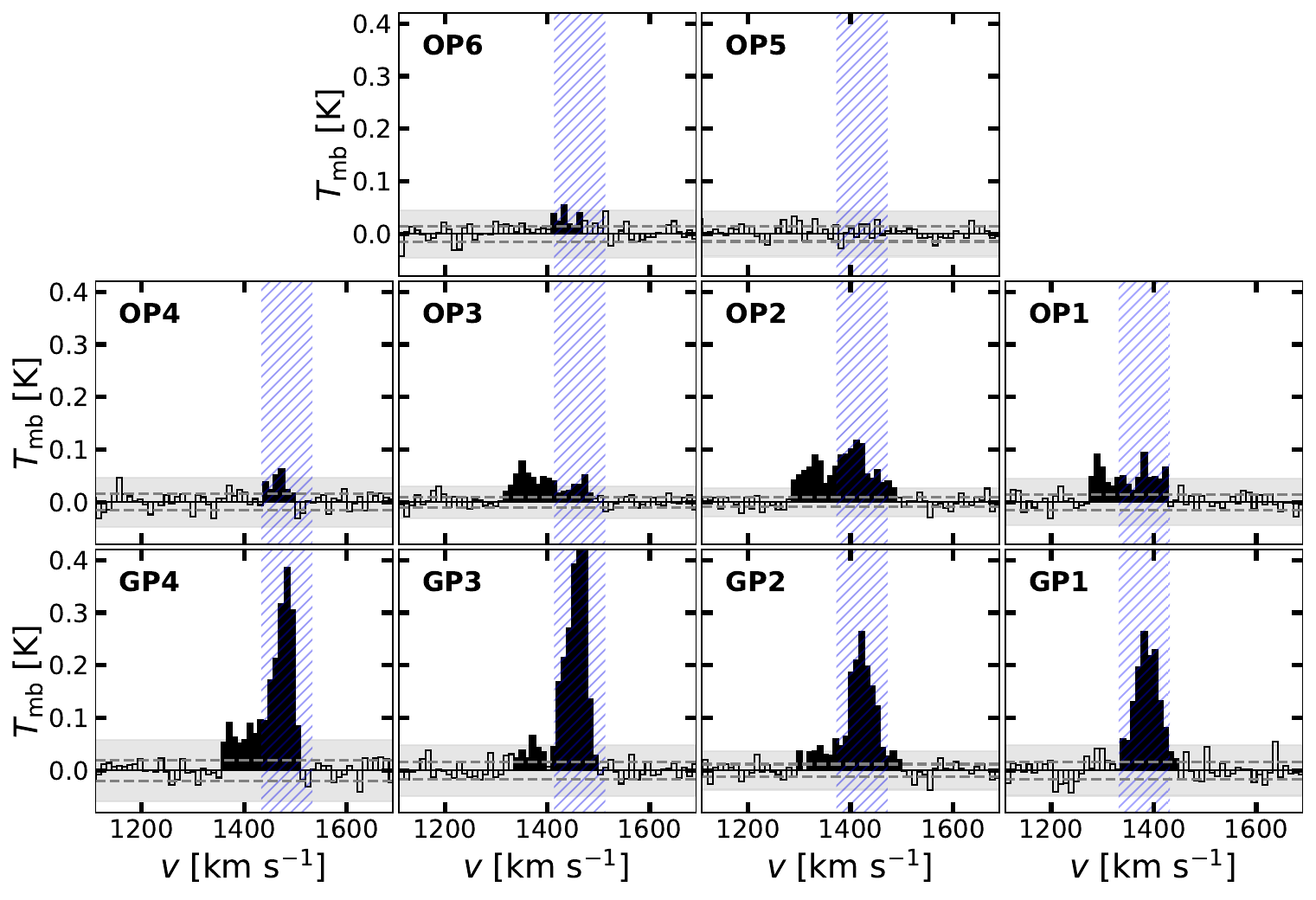} 
     \caption{Spectra of the observed positions along the galactic--plane (GP1--GP4) and off-plane positions (OP1--OP6) in NGC~4565. Each panel shows the $^{12}$CO($J$=1--0) spectrum in 10~km~s$^{-1}$ channels ($T_{\rm mb}$ vs.\ $v$), where all channels are shown as open bars and the channels selected for the $W_{\rm CO}$ integration are highlighted as black filled bars. The gray band indicates the $3\times T_{\rm rms}$ noise level and the dashed horizontal lines correspond to $T_{\rm rms}$. The blue hatched band marks the local disk velocity window, defined as $v_{\rm pk,GP}\pm50~\mathrm{km\,s^{-1}}$, where $v_{\rm pk,GP}$ is the peak velocity of each paired galactic-plane spectrum. The upper left of each subplot shows the position name, corresponding to the definitions in Figure~\ref{fig:fov} and Table~\ref{tab:obs_parameters}.
     {Alt text: Ten CO spectra of galactic-plane and off-plane positions in NGC 4565.}
     }
\label{fig:spectrum}
\end{figure*}

In the following analysis, we integrate the CO emission using a two-threshold scheme to capture faint line wings adjacent to statistically robust detections. We first identify ``core'' channels that satisfy $T_{\rm mb}(v)\ge 3\,T_{\rm rms}$. Starting from each contiguous core segment, we additionally include adjacent ``wing'' channels with $T_{\rm mb}(v)\ge T_{\rm rms}$, allowing up to three consecutive wing channels on each side. We visually inspected all spectra and excluded obvious outliers that are inconsistent with the line profile. For example, the isolated spike at $v\sim1600~\mathrm{km~s^{-1}}$ in GP1 (Figure~\ref{fig:spectrum}) was removed from the analysis. The integrated intensity is computed as
\begin{equation}
W_{\rm CO}\ [\mathrm{K~km~s^{-1}}]
=\sum_{v\in\mathcal{S}} T_{\rm mb}(v)\,\Delta v,
\label{eq:Wco}
\end{equation}
where $\mathcal{S}$ defines the effective integration window determined by the core+wing criterion. In Figure~\ref{fig:spectrum}, the selected channels for the integration are shown as black filled bars, while the remaining channels are shown as open bars. We define an effective line width (equivalent width) as
\begin{equation}
\sigma_{\rm eff}\ [\mathrm{km~s^{-1}}] = \frac{W_{\rm CO}}{T_{\rm pk}},
\label{eq:sigma}
\end{equation}
where $T_{\rm pk}$ is the peak main-beam temperature of the CO spectrum.

We convert $W_{\rm CO}$ to the molecular gas surface density via
\begin{equation}
\Sigma_{\rm mol}\ [M_\odot~\mathrm{pc^{-2}}] = \alpha_{\rm CO}\,W_{\rm CO},
\label{eq:surfacedensity}
\end{equation}
adopting $\alpha_{\rm CO}=4.35~M_\odot~(\mathrm{K\,km\,s^{-1}\,pc^2})^{-1}$, which is commonly used for Milky Way--like environments \citep{BolattoETAL2013}.
The molecular mass per beam is estimated as
\begin{equation}
M_{\rm mol}\ [M_{\odot}] = \Sigma_{\rm mol}\,\Omega_{\rm mb}\,D^2,
\label{eq:mass}
\end{equation}
where $\Omega_{\rm mb} = 1.133\,\theta_{\rm mb}^2$ is the solid angle of the Gaussian main beam (in radians), and $D$ is the source distance. We note that $M_{\rm mol}$ scales linearly with the assumed $\alpha_{\rm CO}$; if $\alpha_{\rm CO}$ differs for off-plane gas compared to the disk, our inferred $M_{\rm mol}$ would change proportionally. In general, $\alpha_{\rm CO}$ is often expected to increase at larger heights above the midplane as the gas becomes more diffuse and CO-dark H$_2$ may be more prevalent. Milky Way measurements toward intermediate/high Galactic latitudes suggest that, despite substantial cloud-to-cloud variations, the mean CO-to-H$_2$ conversion factor can remain comparable to the canonical disk value \citep{ChenETAL2015}. In practice, constraining $\alpha_{\rm CO}$ on a region-by-region basis is challenging, and many studies of nearby galaxies adopt a fixed $\alpha_{\rm CO}$ \citep[e.g.,][]{Garcia-BurilloETAL1992,NagataETAL2025}. We therefore assume a constant $\alpha_{\rm CO}$ in our analysis. We also compare our estimated $M_{\rm mol}$ with values reported in previous studies of other nearby galaxies. A summary of the derived quantities is presented in Table~\ref{tab:obs_parameters}.

\subsection{Off-plane CO emission and basic properties}
\label{subsec:off-emission}
As shown in Figure~\ref{fig:spectrum}, we detect $^{12}$CO($J$=1--0) emission at the off-plane positions OP1--OP4 and OP6 with $T_{\rm pk} \ge 3T_{\rm rms}$, and give the upper limit at OP5. The detections at OP1--OP3 are robust, whereas those at OP4 and OP6 are marginal. The spectra at OP4 and OP6 can be explained by contamination from the disk emission through the telescope beam pattern (see Section~\ref{subsec:beam-pattern} for details). The OP1--3 positions are located above the galactic plane (see Figure~\ref{fig:fov}), corresponding to a projected vertical height of $z_{\rm p}\sim0.85\pm0.2$ kpc. The uncertainty is a conservative upper bound, estimated by propagating the distance error and the possible pointing offset. The molecular gas mass at these off-plane positions (OP1--OP3) is estimated to be $M_{\rm mol}\simeq(2.1$--$4.3)\times10^{7}\,M_\odot$. We also quantify the off-plane molecular fraction by comparing each off-plane position with its corresponding disk position,
\begin{equation}
f_{{\rm off},i}\equiv \frac{W_{{\rm OP},i}}{W_{{\rm GP},i}+W_{{\rm OP},i}},
\end{equation}
where $W_{{\rm OP},i}$ and $W_{{\rm GP},i}$ are the CO integrated intensities ($W_{\rm CO}$) measured at the off-plane (OP$\it{,i}$) and disk (GP$\it{,i}$) positions, respectively. 
The off-plane fractions are 0.35, 0.44, and 0.22 for \(f_{\rm off,1}\), \(f_{\rm off,2}\), and \(f_{\rm off,3}\), respectively. Using the summed intensities over the three positions, we obtain 
$f_{\rm off}\approx 0.34$. This fraction is comparable, within a factor of a few, to the values reported for edge-on galaxy NGC~891 \citep[e.g.,][]{Garcia-BurilloETAL1992,Jimenez-LopezETAL2026arXiv}. This suggests that normal disk galaxies may show thick molecular gas layers, at least locally and at kpc-scale resolution.

To isolate the velocity component associated with local disk rotation, we define a local disk velocity window as $v_{\rm pk,GP}\pm50~\mathrm{km\,s^{-1}}$, where $v_{\rm pk,GP}$ is the peak velocity of the paired GP spectrum. This window is broad enough to include the main local disk component in the GP spectra and is shown by the blue-hatched region in Figure~\ref{fig:spectrum}. Within this window, the off-plane spectra show CO components at velocities close to those of the corresponding galactic-plane positions, indicating that at least part of the off-plane molecular gas shares the rotational motion of the disk. Although weak wing emission is also present in some of the GP spectra, the $\sigma_{\rm eff}$ values of OP1--OP3 range from 83 to 115~km~s$^{-1}$, about twice those measured at the GP positions. The broader off-plane profiles suggest that additional kinematic components contribute more strongly in the OP spectra than in the GP spectra. 

Before interpreting the off-plane emission, we first examine the possible contribution of disk emission entering through the telescope beam pattern in Section~\ref{subsec:beam-pattern}. We then evaluate projection effects arising from the slight deviation of NGC~4565 from an exactly edge-on orientation in Section~\ref{subsec:projection}. Finally, we discuss possible origins of the kinematic differences in Section~\ref{subsec:hv}.

\section{Discussion}\label{sec:discussion}
\subsection{Estimate of beam-pattern effects}\label{subsec:beam-pattern}
The off-plane area may be contaminated by bright disk emission entering through the telescope beam pattern, including sidelobes and broader error-beam components. The FOREST beam pattern reported by \citet{MinamidaniETAL2016} was measured at 86~GHz, whereas the present observations were conducted at 115~GHz. Because no dedicated FOREST beam-pattern measurement is available at 115~GHz, we use the 86~GHz FOREST beam pattern as a reference after scaling its size by the beam FWHMs (i.e. frequencies). This approach is motivated by published Nobeyama 45-m measurements showing broadly similar beam-pattern structures among different receivers when normalized by the FWHM \citep{MinamidaniETAL2016,NakamuraETAL2015,NakamuraETAL2024}. The consistency of these normalized beam patterns over 2015--2024 also suggests that the beam pattern has not changed significantly with time. The scaled 115~GHz beam-pattern indicates that the beam response at one- to two-beam offsets is $\sim$10 to a few $\%$ of the main-beam response. To allow for uncertainty in the 115~GHz response, we conservatively adopt contamination limits of 20$\%$ at one-beam offsets (OP1--OP4) and 10$\%$ at approximately two-beam offsets (OP5--OP6), using the paired GP intensities as references.

Taking the galactic plane $W_{\rm CO}$ as a reference, the observed $W_{\rm CO}$ values (Table~\ref{tab:obs_parameters}) exceed these contamination levels at OP1--3 but not at OP4 and OP6.  Therefore, the detection at OP4 and OP6 can be explained by conservative beam-pattern contamination, and we exclude these two positions from the following discussion.

\subsection{Vertical structure of molecular gas accounting for projection effects}\label{subsec:projection}

A potential alternative explanation for the apparent off--plane CO emission is a purely geometrical one, in which an intrinsically thin molecular disk viewed at a slight deviation from perfectly edge-on ($i\simeq88.5^\circ$, with $i=0^\circ$ face-on) is broadened by projection. To evaluate this possibility, we first quantified how beam convolution set the \emph{apparent} vertical extent. Given the Nobeyama-45\,m beam size (${\rm FWHM}\simeq14\arcsec\simeq0.8$~kpc), even a completely edge-on razor-thin disk would appear broadened by roughly one beam width after convolution.

Most simply, the projection--induced thickness of a razor-thin disk depends on the galactocentric radius ($R_{\rm gal}$), and the apparent full thickness produced purely by projection is
\begin{equation}
    h_{\rm geo}(R_{\rm gal})
    =
    2\sqrt{R_{\rm mol}^{2}-R_{\rm gal}^{2}}\,
    \sin (90^{\circ}-i),
\end{equation}
where \(R_{\rm mol}\) is the molecular disk radius and \(i\) is the inclination. Adopting \(R_{\rm mol}\simeq0.7R_{25}\) \citep{CormierETAL2016}, we obtain \(R_{\rm mol}\simeq19.3\) kpc and \(h_{\rm geo}(0)\simeq1.0\) kpc at the galactic centre of NGC~4565. In the observed field, where \(R_{\rm gal}\simeq4.5\) kpc, the same estimate gives \(h_{\rm geo}(4.5~{\rm kpc})\simeq0.98\) kpc. We note that \(h_{\rm geo}\) represents the full projected thickness; the corresponding one-sided projected extent is \(h_{\rm geo}/2\sim0.49\) kpc at the observed regions.

The OP1--OP3 beams are centred at $z_{\rm p}\sim0.85$ kpc from the GP positions, which is larger than the one-sided projected extent of a razor-thin disk for the fiducial inclination of \(i=88.5^\circ\). With the lower inclination of \(i=87.5^\circ\) \citep{ZschaechnerETAL2012}, \(h_{\rm geo}/2\) becomes \(\simeq0.82\) kpc in the observed field, comparable to the OP1--OP3 heights. Therefore, projection effects cannot be ruled out by this simple estimate alone. To discuss this possibility more quantitatively, we perform the forward modelling as described below.

We examined how large a vertical thickness of the molecular gas disk is required to reproduce the observed $f_{\rm off}$ values. We constructed simplified forward models using the \textsc{GALMOD} task in \textsc{3D-Barolo} \citep{Teodoro&Fraternali2015}. The model represents the CO disk as a set of concentric, axisymmetric tilted rings. We explicitly fix the basic geometry and kinematics (i.e., the dynamical centre, systemic velocity, inclination, and position angle; see Table~\ref{tab:ngc4565_parameters}) and adopt a constant rotation velocity at all radii together with a constant velocity dispersion of $10~\rm km\,s^{-1}$ \citep{Caldu-PrimoETAL2013,KodaETAL2023}. To examine the effect of a lower inclination, we also tested models with \(i=88.0^\circ\) and \(87.5^\circ\), in addition to the fiducial \(i=88.5^\circ\) model. We also tested models with velocity dispersions of 5, 10, 15, and  20~km~s$^{-1}$, and confirmed that this choice does not significantly affect the following discussion. The radial surface-brightness distribution is prescribed as an exponential disk profile, \(I(R)\propto \exp(-R/R_{\rm d})\), where we adopt \(R_{\rm d}=0.3R_{\rm mol}\). The vertical gas distribution is described by a Gaussian profile with scale height \(z_0\). We did not consider spiral structure, flares, or warps. We systematically varied the vertical scale height over \(z_0=0\)--2.0~kpc (in 50~pc steps), generated synthetic mock-galaxy cubes on a high-resolution grid (\(1\arcsec\) pixels and 2.5~km\,s\(^{-1}\) channels), and extracted mock spectra at the same GP and OP positions after applying a \(14\arcsec\) beam, matching the angular and velocity resolution of the Nobeyama 45-m observations. Additional tests with model beam patterns consistent with our conservative estimates of beam-pattern contamination (Section \ref{subsec:beam-pattern}) showed that changes in \(f_{\rm off}\) are less than 0.01.

\begin{figure}
  \includegraphics[width=\linewidth,trim={5 0 0 5}, clip]{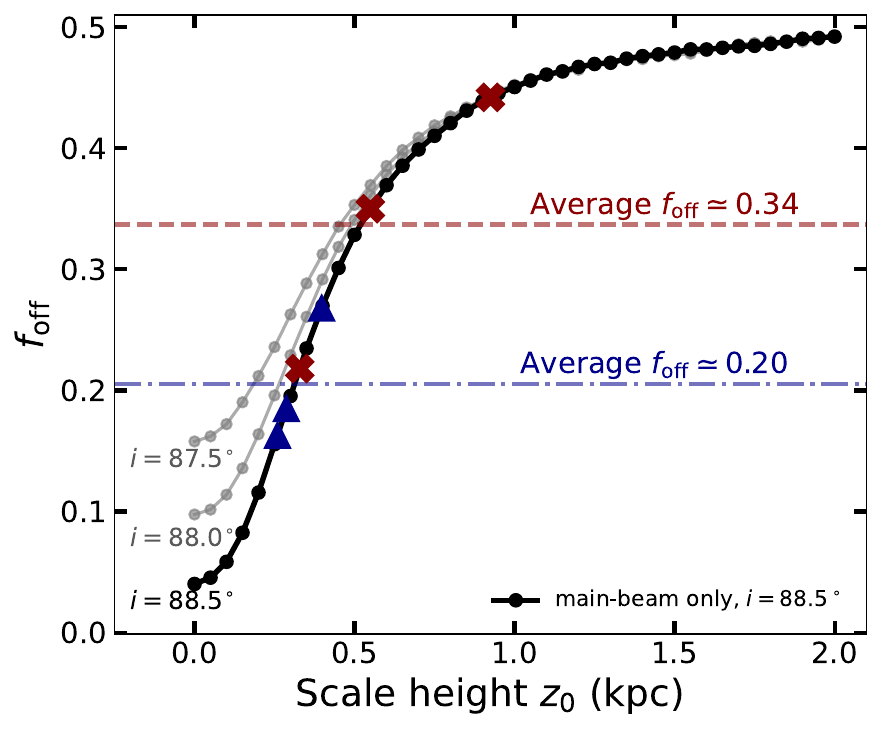} 
    \caption{Comparison between the observed off-plane molecular gas fraction and the results from mock observations of the 3D-Barolo \textsc{GALMOD} task. The black curve shows the fiducial model with \(i=88.5^\circ\). The red crosses show \(f_{\rm off}\) integrated over the full velocity range, while the blue triangles indicate \(f_{\rm off}\) computed only within the hatched velocity interval in Figure~\ref{fig:spectrum}. For both the red crosses and blue triangles, the data points correspond to OP3, OP1, and OP2 from lower to higher \(f_{\rm off}\). We note that the symbols do not represent independent measurements of $z_0$; their horizontal positions indicate the scale heights inferred by matching the observed $f_{\rm off}$ values to the adopted $i=88.5^\circ$ model. The horizontal red dashed and blue dash-dotted lines denote the mean \(f_{\rm off}\) values of the red crosses and blue triangles, respectively.  The gray curves show model predictions for \(i=87.5^\circ\) and \(i=88.0^\circ\), in addition to the fiducial \(i=88.5^\circ\) model.
    {Alt text: Off-plane CO fraction versus molecular-gas scale height.}
    }
\label{fig:foff}
\end{figure}

Figure~\ref{fig:foff} shows how \(f_{\rm off}\) depends on the scale height due to intrinsic thickness of the disk. As \(z_0\) increases, \(f_{\rm off}\) rises monotonically. In the model galaxy, we confirmed that the variation in \(f_{\rm off}\) among the observed position pairs  (OP1/GP1--OP3/GP3) is small, and thus adopted the mean value as the observed value. Figure~\ref{fig:foff} also shows models with different inclinations, \(i=87.5^{\circ}\), \(88.0^{\circ}\), and \(88.5^{\circ}\). The inclination affects the model prediction mainly at small \(f_{\rm off}\), while the curves become similar in the range \(f_{\rm off}\gtrsim0.34\), where the observed off-plane fractions are located.

A large-scale warp could also affect the apparent vertical distribution. However, \citet{Martinez-LombillaETAL2023} show that the warp becomes important mainly in the outer disk at \(R_{\rm gal}\gtrsim20\)~kpc in this galaxy. This radius is larger than the adopted CO disk radius, \(R_{\rm mol}=0.7R_{25}\simeq19.3\)~kpc. This places the prominent large-scale warp mostly outside the CO-emitting disk considered here, making a large-scale warp unlikely to be the primary origin of the observed high \(f_{\rm off}\) in this field.

For the observational measurements, the red crosses show \(f_{\rm off}\) computed by integrating over the full velocity range  (i.e., all detected velocity components), and the blue triangles show \(f_{\rm off}\) computed using only the velocity components highlighted by the blue-hatched interval. Only the vertical positions of these symbols represent the observed \(f_{\rm off}\) values; their horizontal positions are placed at the intersections with the adopted \(i=88.5^\circ\) model curve and indicate the scale height \(z_0\) required to reproduce each observed fraction. The observed mean value, shown by the horizontal red dashed line, corresponds to \(z_0\sim0.5\)~kpc in this model. Moreover, reproducing the largest observed value, \(f_{\rm off,2}=0.44\), within this inclined disk model requires a scale height of \(\sim0.9\)~kpc, much larger than expected for a canonical thin molecular disk. These results imply that the observed off-plane emission cannot be explained solely by a geometrically thin disk affected by projection effects. The \(i=87.5^\circ\) model slightly increases the projected disk contribution, but still cannot reproduce the observed full-velocity \(f_{\rm off}\). Instead, reproducing the observed \(f_{\rm off}\) requires molecular gas to be present at kpc-scale heights above the disk in this region.

Figure~\ref{fig:foff} also highlights that $f_{\rm off}$ depends strongly on the adopted velocity range. When the calculation is restricted to the blue-hatched interval, i.e., the velocities consistent with disk rotation, the mean off-plane fraction is reduced to \(f_{\rm off}\simeq0.2\). The large difference arises because OP1--OP3 show velocity components offset from \(v_{\rm pk,GP}\). We discuss the nature and implications of these velocity-offset components in more detail in Section~\ref{subsec:hv}.

\subsection{High-velocity component in the off-plane region}\label{subsec:hv}
While the off-plane molecular gas might reflect locally thick disk components, we here examine an additional origin of this emission. As shown in Figure~\ref{fig:spectrum}, the off-plane positions OP1--OP3 exhibit large $\sigma_{\rm eff}$ values that partly reflect the presence of two velocity components. One component is close to the local disk velocity, while the other is offset from it by $\Delta V\sim100~\mathrm{km\,s^{-1}}$. Given the lack of comparable high-velocity emission in the paired GP1--GP3 spectra, this component is unlikely to be simply the same local disk component. We hereafter refer to the emission outside the blue-hatched local disk velocity window as an observationally defined high-velocity component. We estimate $M_{\rm mol}^{\rm HV}$ from the CO integrated intensity outside the blue-hatched local disk velocity window in Figure~\ref{fig:spectrum}, using the same $\alpha_{\rm CO}$ as adopted above. We obtain $M_{\rm mol}^{\rm HV}\simeq (1.1$--$2.0)\times10^{7}\,M_\odot$ at each OP position.

To assess whether stellar feedback energy from the galactic disk could accelerate this molecular gas, we estimate the kinetic energy associated with the component following \citet{NagataETAL2025}:
\begin{equation}
    E_{\rm kin} = \frac{1}{2}\, M_{\rm mol}^{\rm HV}\,\Delta V^2 .
\end{equation}
Adopting a characteristic relative velocity of $\Delta V \simeq 100~\mathrm{km\,s^{-1}}$ and $M_{\rm mol}^{\rm HV}\simeq (1.1$--$2.0)\times10^{7}\,M_\odot$, and assuming that the entire high-velocity molecular gas component moves at this velocity, we derive $E_{\rm kin}\simeq (1.1$--$2.0)\times 10^{54}\,\mathrm{erg}$. This value is a lower limit, because neither a de-projection correction for the inclination of NGC~4565 nor the energy required to lift the gas against the stellar disk gravity is included. If only a small fraction ($\lesssim 10\%$) of a typical supernova explosion energy \citep[$\sim 10^{51}\,\mathrm{erg}$;][]{Chevalier1974} couples to the ISM, the implied energy budget corresponds to $\sim10^{4}$ supernovae within a single $\sim1\,\mathrm{kpc}$ beam. Given the Milky-Way-like global SFR of NGC~4565, concentrating such a large amount of feedback energy within a $\sim1$ kpc region would require intense local star formation. This makes a disk-driven outflow or fountain an unlikely origin for the observed high-velocity gas. However, because NGC~4565 is highly inclined but not exactly edge-on, a projected spiral-arm crossing could also contribute to this velocity-offset emission. We therefore regard external inflow as one possible origin, rather than a unique interpretation.

A similar interpretation has been proposed for nearby face-on galaxy M83, which is also a Milky-Way analogue in terms of its global properties \citep[see][for a summary]{KodaETAL2023}. In this galaxy, \citet{NagataETAL2025} reported ten molecular HVCs ($\Delta V\gtrsim 50~\mathrm{km\,s^{-1}}$) and interpreted them as accreting material based on their kinematics and energetics. Several regions host three to four such clouds clustered within $\lesssim$1~kpc--comparable to the effective spatial scale of our measurements. The total molecular mass of HVCs in M83 is $\sim10^{7}\,M_\odot$, in good agreement with the mass derived in this study. 

In the Milky Way, CO emission from high-velocity clouds is generally elusive \citep[e.g., non-detections in $^{12}$CO~($J$=1--0);][]{Dessauges-ZavadskyETAL2007}. Although those studies reached even better sensitivity than ours, the number of targeted HVC sightlines remains small, and conducting an all-sky survey at comparable depth is extremely challenging. Wide-field, high-sensitivity molecular-gas mapping of nearby edge-on galaxies covering both the full disk and off-plane regions, therefore, offers a feasible route to distinguish external inflow from projected spiral-arm crossing and to characterize molecular gas well above galactic disks and to constrain the physical conditions of such high-velocity components. In this context, recent and ongoing survey projects such as CHANG-ES \citep[Continuum Halos in Nearby Galaxies: An EVLA Survey;][]{IrwinETAL2012} and GECKOS \citep[Generalising Edge-on galaxies and their Chemical bimodalities, Kinematics, and Outflows out to Solar environments;][]{van-de-SandeETAL2024} provide valuable reference and comparison datasets and may play a critical role in elucidating the origin of off-plane molecular gas in galaxies.

\section{Conclusions}\label{sec:conclusion}
Using the Nobeyama 45-m telescope, we detected significant $^{12}$CO ($J$=1--0) emission above the disk of the Milky Way--like galaxy NGC~4565 at an angular resolution of 14\arcsec, corresponding to $\sim0.8$ kpc. After evaluating possible beam-pattern contamination, three off-plane positions (OP1--OP3) show robust CO detections with molecular masses of $(2.1$--$4.3)\times10^7\,M_{\odot}$. Even after accounting for geometric projection effects, the observed off-plane emission cannot be explained solely by a thin-disk model, suggesting that at least part of the emission arises from molecular gas located above the disk. We also identify a high-velocity component ($\Delta V\sim100~\rm km~s^{-1}$) with a mass of $(1.1$--$2.0)\times10^7\,M_{\odot}$; if this component is not dominated by projected disk emission, its kinetic-energy suggests the external inflow as its origin.

These results provide observational evidence that molecular gas can form a vertically extended component at kpc heights even in a non-starburst disk, highlighting a possible gas-replenishment channel in the disk-halo interface. However, our conclusions are limited to the filament-targeted fields and do not establish a galaxy-wide thick molecular disk; uniform wide-field CO surveys are required to assess the ubiquity of such components and to quantify their vertical structure, physical conditions, and role in the disk--halo ISM cycle.


\par\noindent
\begin{ack}
We thank the anonymous referee for constructive comments that improved the manuscript. We are grateful to Dr. Atsushi Nishimura for his assistance during the observations. We also thank Kotaro Kohno, Jin Koda, and Maki Nagata for helpful discussions. The Nobeyama 45-m radio telescope is operated by the Nobeyama Radio Observatory, a branch of the National Astronomical Observatory of Japan (NAOJ). Data analysis was in part carried out on the Multi-wavelength Data Analysis System operated by the Astronomy Data Center (ADC), NAOJ. F.M. is supported by JSPS KAKENHI grant Nos. JP23K13142 and JP23K20035.
\end{ack}


\par\noindent
\section*{Data availability}
The raw Nobeyama 45-m data underlying this study will be released through the Nobeyama Science Archive (\url{https://nobeyama-archive.nao.ac.jp/}) after the proprietary period. Mock data used for testing and visualization were generated with the 3D-Barolo algorithm (\texttt{pyBBarolo v1.3}), which is publicly available at \url{https://editeodoro.github.io/Bbarolo/}.

\bibliographystyle{apj}
\bibliography{reference}

\end{document}